\documentclass[preprint,superscriptaddress,pre,floats,aps,amsmath,amssymb,12pt]{revtex4-1}

\usepackage{enumerate,ulem,graphicx,array}
\usepackage[english]{babel}
\usepackage[utf8]{inputenc}
\usepackage[T1]{fontenc}
\usepackage[left=1.5cm, right=1.5cm, top=1.785cm, bottom=2.0cm]{geometry}
\usepackage{balance}
\usepackage{times,mathptmx}
\usepackage{lastpage}
\usepackage{float}
\usepackage{fnpos}

\usepackage[usenames,dvipsnames]{xcolor}
\usepackage{setspace}
\usepackage{hyperref}

\begin{document}

\title{Speciation of FeS and FeS$_2$ by means of X-ray Emission Spectroscopy using a compact full-cylinder von Hamos spectrometer}

\author{Malte Wansleben}
\altaffiliation[Present address: Helmut Fischer GmbH, R\"{o}ntgen- und Neutronenoptiken, Rudower Chaussee 29/31, 12489 Berlin, Germany. E-mail: ]{malte.wansleben@helmut-fischer.com}
\affiliation{Physikalisch-Technische Bundesanstalt, Abbestrasse 2-12, 10587 Berlin, Germany.}
\affiliation{Helmholtz-Zentrum Berlin f\"{u}r Materialien und Energie GmbH, Hahn-Meitner-Platz 1, 14109 Berlin, Germany.}
\author{John Vinson}
\affiliation{National Institute of Standards and Technology, Gaithersburg, MD 20899, USA.}
\author{Andr\'{e} W\"{a}hlisch}
\author{Karina Bzheumikhova}
\author{Philipp H\"{o}nicke}
\author{Burkhard Beckhoff}
\author{Yves Kayser}
\email[E-mail: ]{yves.kayser@ptb.de}
\affiliation{Physikalisch-Technische Bundesanstalt, Abbestrasse 2-12, 10587 Berlin, Germany.}

\begin{abstract}
We present Fe K\(\beta\) X-ray emission (XES) and Fe K X-ray absorption spectra (XAS) of Iron(II)sulfide (FeS) and Iron(II)disulfide (FeS$_2$). While XES and XAS offer different discrimination capabilities for chemical speciation, depending on the valence states of the compounds probed, XES allows for using different excitation sources. The XES data was measured using polychromatic X-ray radiation with a full-cylinder von Hamos spectrometer being characterized by an energy window of up to 700 eV and a spectral resolving power of $E/ \Delta E = 800$. The large energy window at a single position of the spectrometer components is made profit of to circumvent the instrumental sensitivity of wavelength-dispersive spectrometers to sample positioning. This results in a robust energy scale which is used to compare experimental data with {\it ab initio} valence-to-core calculations, which are carried out using the {\sc ocean} package. To validate the reliability of the {\sc ocean} package for the two sample systems, near edge X-ray absorption fine structure measurements of the Fe K absorption edge are compared to theory using the same input parameters as in the case of the X-ray emission calculations. Based on the example of iron sulfide compounds, the combination of XES experiments and {\sc ocean} calculations allows unravelling the electronic structure of different transition metal sulfides and qualifying XES investigations for the speciation of different compounds. 
\end{abstract}

\maketitle

\section{Introduction}
With renewable energies on the rise and the ongoing change from combustion based transportation to electric transportation there is a growing demand for electric energy storage devices. In this respect metal sulphides are cheap, abundant, and environmental friendly materials for energy storage applications \cite{C7RA03599C, doi:10.1002/aenm.201703259, Zhao2018} and chemistry\cite{doi:10.1080/01614948408064718, life8040046, MORALESGALLARDO201693, doi:10.1021/acs.accounts.7b00187, doi:10.1021/jp506288w}.

More specifically, pyrite (FeS$_2$) has been demonstrated to be of use in batteries\cite{ZHAO201762} and remains a compound of interest for improving the cycling performance of Li/S batteries. Furthermore, iron sulfide species and the discrimination between them are of importance in chemical reactions and processes\cite{C6CP03760G, C7CC00120G, Thiel6897}. In this manuscript we investigate the iron sulfur compounds Iron(II)sulfide (troilite, FeS) and Iron(II)disulfide (pyrite, FeS$_2$) using X-ray emission spectroscopy (XES) in the tender X-ray range by means of a wavelength-dispersive spectrometer based on a modified von Hamos geometry. These two simple iron sulfides have been subject to research, experimentally and theoretically, for many years with respect to their band structure in order to understand electrical, magnetic and optical properties\cite{Joe2016, Mortensen2017, Lauer1984, Folkertst1987, Fourcadetj1997, Tossell1977, Antonov2009,RAO1976207}. We demonstrate the discrimination capability of theoretical and experimental methods for the chemical speciation and therewith provide benchmark results for the Fe K XES and X-ray absorption fine structure (XAFS) spectroscopy. While XAFS experiments require incident radiation with a tunable photon energy as offered commonly at synchrotron radiation beamlines, XES experiments can be realized also with radiation sources having a broad energy bandwidth, e.g. non-monochromatized synchrotron radiation (white light) or Bremsstrahlung from an X-ray tube (laboratory instrumentation). 

The study of K$\beta$ X-ray emission spectra of 3d transition metal compounds by means of high-resolution XES is a suitable and established experimental tool in a broad range of research fields that aim to gain information on the valence electronic structure \cite{doi:10.1021/ac900970z, Kowalska2015, C4CP00904E}. The energies of these chemically sensitive K$\beta$ emission lines lie in the tender X-ray regime, ranging from 4.0 keV to 9.5 keV, ensuring large penetration depths and flexibility with respect to the sample environment, i.e. favourable conditions for in-situ and operando experiments, as opposed to the L-edge spectroscopy which, however, provides a direct access to the partially filled 3d states via dipole transitions.

In XES the radiative relaxation (fluorescence emission) of a core hole is analyzed \cite{Bergmann2009}. The K fluorescence emission, which results from relaxation processes following an ionization of the atom in the K shell, is composed of two main regimes caused by transitions of electrons of principle quantum number $n=2$ (K\(\alpha\)) and principle quantum numbers $n=3,4$ (K\(\beta\)). Due to spin and angular momentum the transitions are further subdivided into K\(\alpha_1\) and K\(\alpha_2\), as well as K\(\beta_{1,3}\) and K\(\beta_{2,5}\), respectively. The K\(\beta\) emission lines exhibit a more distinct sensitivity to the electronic structure of the partially filled 3d electron shell and other occupied valence states that are involved in chemical bonding\cite{Kowalska2015}. While the modification of the K\(\beta_{1,3}\) main line is dominated by spin state contributions\cite{Vanko2006,Joe2016}, the spectral region around the K\(\beta_{2,5}\) line is more sensitive to changes in the valence state energies\cite{Bergmann1999, Mortensen2017, Wansleben2018}. The spectroscopy of this spectral region is referred to as valence-to-core XES (vtc-XES)\cite{Gallo2014}.

As the features associated with vtc-XES exhibit low transition probabilities as well as small energy differences on the order of a few eV or less, measurements require instrumentation capable of efficiently resolving these features. There are many spectrometer types that meet this requirement, ranging from  transition edge sensors ($E/\Delta E =$  1300)\cite{Joe2016} to multi-crystal wavelength-dispersive spectrometers (WDS, $E/\Delta E =$ 5000 to 10000) with spatial dimensions exceeding several meters\cite{Sokaras2013, Alonso-Mori2012a}. There are generally three characteristics that classify these spectrometers: energetic resolving power, detection efficiency, and energy window defining the accessible spectral energy range at a preselected, fixed spectrometer setting\cite{Hoszowska1996}. While the resolving power is usually the most interesting parameter as it is the significant number for revealing details in the experimental data, efficiency and energy window are more of practical nature. Higher efficiency enables shorter measurement times, lower detection limits and possibly circumvents radiation damage to the sample. A large energy window allows covering a broad spectral region enabling the analysis of widely distributed spectroscopic features with only one spectrometer setting, e.g., simultaneous detection of 3d metal K$\alpha$ and K$\beta$ emission lines without moving any spectrometer component during which one of the key parameters of a spectrometer could be modified. 

The XES studies were done using undispersed synchrotron radiation of a bending magnet. The X-ray beam was collimated by means of a pinhole and a von Hamos spectrometer based on a full-cylinder Bragg crystal with a small radius of curvature. While the full cylinder geometry in conjunction with the use of mosaic crystals allows to achieve an optimized detection efficiency, the small radius of curvature permits covering an larger energy window at a given crystal width and a fixed position of the spectrometer components. The compromise made by selecting the small radius of curvature is a reduced resolving power of the spectrometer. We demonstrate that the resolving power achieved by the applied spectrometer is sufficient to realize chemical speciation of a 3d transition metal by means of K$\beta$ spectroscopy including vtc-XES. Additionally, the spectrometer's energy window of over 700~eV allows for referencing the chemical sensitive K$\beta$ line energies to the K$\alpha$ doublet. This alleviates the challenge of correcting the energy scale to account for possible changes in the sample position when using different reference lines. The sensitivity of WDS to sample position commonly requires employing entrances slits, resulting in a potential loss of detection efficiency\cite{Mortensen2016}. 

The presented XES data is complemented with {\it ab initio} calculations using the {\sc ocean} package\cite{Vinson2011,Gilmore2016}. Furthermore, we compare near-edge X-ray absorption fine structure spectra (NEXAFS) of the two compounds with calculations using the same input parameters as for XES to validate the theory and demonstrate the performance of the {\sc ocean} package in this X-ray regime. A recent publication has already demonstrated the potential in off-resonant XES by validating vtc-XES of titanium oxides\cite{Wansleben2018}.

\section{Calculation}
Calculations of XES and XAFS were carried out using the {\sc ocean} package\cite{Vinson2011,Gilmore2016}. 
Electron orbitals were calculated using density-functional theory as implemented in the {\sc Quantum ESPRESSO} code\cite{espresso}, using the generalized gradient approximation \cite{PhysRevLett.77.3865}. Pseudopotentials were taken from the PseudoDojo collection \cite{pspdojo1,PhysRevB.88.085117}. An energy cut-off of 120~Ry was used for both materials. Within {\sc ocean}, absorption spectra are calculated using the Bethe-Salpeter equation, requiring electron orbitals for both the final states and for a calculation of the core hole screening. For FeS$_2$ the electron orbitals were calculated on an 8$^3$ shifted k-point mesh with 212 conduction band states for the final states, while the screening was carried out using a 2$^3$ k-point mesh and 412 conduction bands. The parameters for FeS were scaled by the differences between the two unit cell sizes with a $6\times6\times4$ k-point mesh and 484 conduction bands for the final states and a $2\times2\times1$ k-point mesh and 940 conduction bands for the screening. Experimental lattice constants were used for both FeS$_2$\cite{FeS2cell} and FeS\cite{Wyckoff}. Both dipole and quadrupole terms are included in the transition operator for absorption and emission. In addition to 0.6~eV broadening to simulate the Fe 1s core-hole lifetime, Gaussian broadening has been applied to the calculated XAS spectra and Lorentzian broadening has been applied to the calculated vtc-XES spectra to match experiment. 
Note that the calculation of XES spectra is focused on the vtc-XES region of the K$\beta_{2,5}$ line because there are some challenges in describing the 3p-3d splitting which results in the occurrence of the K$\beta'$ satellite line \cite{Mortensen2017}. The energy of the core level is missing in {\sc ocean} calculations. 
A single offset parameter is required to shift the calculated spectra to match experiment \cite{PhysRevB.85.045101}. The same parameter is used for both the Fe K absorption and emission and shared between FeS$_2$ and FeS.

\section{Experiment}
All measurements were realized in the Physikalisch-Technische Bundesanstalt laboratory at the electron storage ring BESSY II\cite{Beckhoff2009}. The XAFS spectra were measured at the four-crystal-monochromator (FCM) beamline\cite{Krumrey1998,Krumrey2001} in transmission mode using a photodiode as detector and tuning the incident photon energy over the Fe K absorption edge, referred to as near-edge X-ray absorption fine structure (NEXAFS). For each incident photon energy the detected photocurrent was averaged over a five time readout of the photodiode. The XES measurements were performed at the dipole-white light beamline\cite{Thornagel2001} in the same laboratory delivering a polychromatic excitation spectrum corresponding to bending magnet radiation with a critical energy of 2.5 keV originating from a 1.3 Tesla dipole magnet. A beamline filter of 2~$\mu$m aluminum is used to attenuate the polychromatic excitation spectrum below the Fe K-edge and, thus, to drastically reduce unwanted background radiation caused by scattering.

The von Hamos spectrometer was built in-house and can be operated in a single or double Bragg crystal configuration \cite{Holfelder2018}. The spectrometer is attached to an ultra-high vacuum chamber that contains a sample manipulator and a photo diode for transmission measurements similar to the chamber presented by Lubeck at al.\cite{Lubeck2013}. A position sensitive back?illuminated in?vacuum X?ray charge-coupled device (CCD) with 2048~pixels $\times$ 2048~pixels and 13.5~$\mu$m $\times$ 13.5~$\mu$m pixel size is used as the detector unit. The diffractive element consists of a 40-$\mu m$-thick highly-annealed pyrolytic graphite (HAPG) mosaic crystal\cite{Legall2006, Grigorieva2019} mounted on the inner lateral surface of a Zerodur cylinder with a radius of only 50~mm. This makes this device very compact such that it can readily be used at different synchrotron radiation beamlines or with laboratory X-ray sources. HAPG exhibits high reflectivity\cite{Gerlach2015}, and it can be mounted to almost any shape suitable for X-ray optics. In the double crystal configuration, the spectrometer was already successfully applied to vtc-XES of binary titanium compounds, achieving a resolving power of $E/\Delta E$ = 2700\cite{Wansleben2018}. The resolving power of a von Hamos spectrometer generally depends on the radius of curvature. Previous publications of Anklamm et al.\cite{Anklamm2014} and Malzer et al.\cite{Malzer2018} report on full-cylinder HAPG based von Hamos spectrometer with radius of curvature of 150~mm ($E/\Delta E$ = 2000) and 300~mm ($E/\Delta E$ = 4000), scaling the entire setup to multiple meters and, hence, dedicating these devices to stationary use only. A fit of the measured K$\beta_{1,3}$ line of FeS$_2$ and taking its natural line width of 3.53~eV\cite{Holzer1997} into account, the achieved resolving power of the used setup is in the order of $E/\Delta E = 800$ which corresponds to roughly 9~eV full width at half maximum (FWHM). Furthermore, the ratio of cylinder height to cylinder radius defines the energy window of a von Hamos spectrometer. In order to maintain the same energy window per crystal position, increasing the radius results in a quadratic increase of the overall needed crystal surface and therewith material. Thus, the advantage of a small radius of curvature in this regard becomes obvious. The employed cylinder has a width of 20~mm along the cylinder axis such that it enables covering a very large energy window of up to 700~eV. As the spectral resolving power depends on the source size, the excitation beam is focused by means of a polycapillary X-ray lens achieving a source size of around 60 $\mu$m. 

Two different CCD and HAPG crystal positions were used to measure the full Fe K emission spectrum and the Fe K$\beta$ emission lines only. 
The K$\beta$ XES spectra are obtained by summing up 220 CCD images of 40~seconds exposure time each, resulting in a total measurement time of 2h 26 min per spectrum. The spectrum showing Fe K$\alpha$ and Fe K$\beta$ emission to reference the chemically sensitive energy of the K$\beta$ lines is derived from only one CCD image of 40~seconds exposure time because the transition probabilities of these main lines are rather high. Both samples were prepared from high-purity powders as powder spread on adhesive Kapton tape. The samples were set to an incident and emission angle of 45$^{\circ}$. 

\section{Emission spectroscopy}
\begin{figure}
	\begin{center}
	\includegraphics[width=0.85\textwidth]{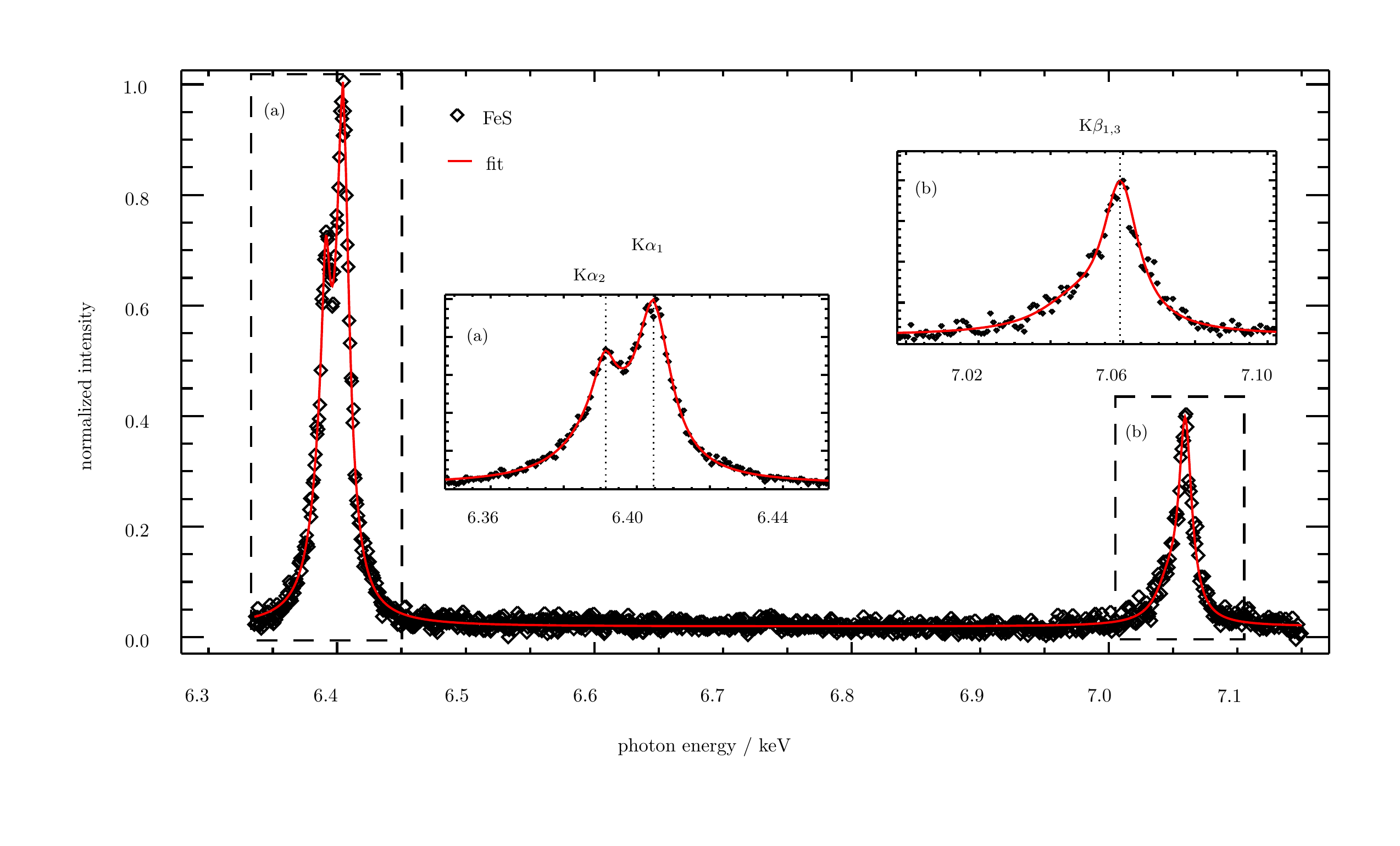}
	\end{center} 
	\caption{Fe K\(\alpha\) and K\(\beta\) emission lines of FeS demonstrating the large energy window covered by the spectrometer. Assuming a negligible chemical shift of K$\alpha_1$, 6404.01(1)~eV\cite{Holzer1997}, and K$\alpha_2$, 6391.03(1)~eV\cite{Holzer1997}, the K$\beta_{1,3}$ line energy is determined to be 7059.4(10)~eV. The experimental uncertainty consists of the uncertainty caused by the polar integration of the CCD images (0.5 eV) and the non-linear least square fitting procedure (0.5 eV). The two contributions are summed up leading to a more conservative uncertainty estimation as compared to the square root of the quadratic sum\cite{Wansleben_Gd}. Relative line intensities of K\(\alpha\) and K\(\beta\) are influenced by the experimental setting and are not corrected in the displayed spectrum. The exposure time was 40~seconds.}
	\label{pic:FeS_Ka_Kb}
\end{figure}

The energy scale of the K$\beta$ XES spectra is referenced to the Fe K$\alpha_1$ and K$\alpha_2$ assuming a negligible chemically induced energy shift of the doublet. The works of Glatzel et al.\cite{Glatzel2005} and Joe et al.\cite{Joe2016} report on a chemical sensitivity of the K$\alpha$ emission with regard to the line width and relative transition probability of the K$\alpha_1$ and K$\alpha_2$ emission lines. However, only small energetic shifts on the order of less than 1~eV of the respective centroid are expected. Hence, by referencing the K$\beta$ emission lines to the K$\alpha$ emission lines we omit the sensitivity of the spectrometer energy scale to the sample positioning. Figure \ref{pic:FeS_Ka_Kb} displays the complete Fe K emission of FeS and illustrates the large energy window covered by the spectrometer in the single crystal configuration. The photon energies and uncertainties of K$\alpha_{1,2}$ emission lines are taken from Holzer et al.\cite{Holzer1997}. Radius channel and photon energy are geometrically related by the Bragg equation as explained in Anklamm et al.\cite{Anklamm2014}. Note that the resolving power of the spectrometer depends on the radii of the detected circles \cite{Wansleben_Gd}.

\begin{figure}
	\begin{center}
	\includegraphics[width=0.85\textwidth]{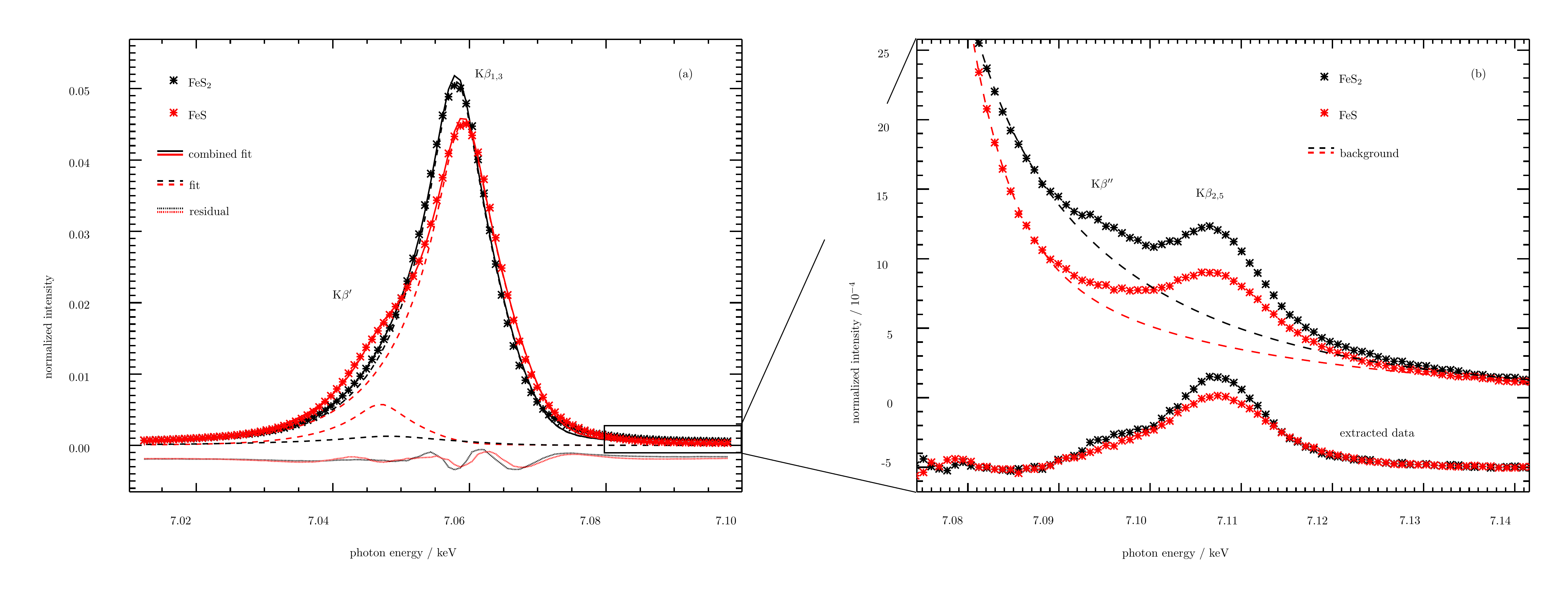}
	\end{center} 
	\caption{(a) Fe K$\beta_{1,3}$ and K$\beta'$ lines and (b) the magnified vtc-XES of FeS$_2$ (black) and FeS (red). The spectra in (a) are fitted with an asymmetric Pseudo-Voigt function \cite{Schmid2014}. The extracted data in (b) is shifted for better visibility. The exposure time was 2h~26min.}
	\label{pic:FeS_FeS2_Kbeta}
\end{figure}

The Fe K$\beta$ emission spectra of FeS and FeS$_2$ are displayed in Figure \ref{pic:FeS_FeS2_Kbeta} (a). The spectra are normalized to their respective area and fitted with an asymmetric Pseudo-Voigt function \cite{Schmid2014}. A shift of the K$\beta_{1,3}$ line to higher photon energies (1.7(5)~eV) and a K$\beta'$ satellite on the low-energy side ($E_{K\beta_{1,3}} - E_{K\beta '}$ = 12(1)~eV) for FeS as compared to FeS$_2$ is recorded. The respective uncertainties of the energy shifts are estimated as described in Wansleben et al\cite{Wansleben_Gd}. The K$\beta'$ satellite is caused by the final state exchange interaction between the 3p hole and 3d valence states allowing for conclusions on the spin state (e.g. number of unpaired electrons)\cite{Rovezzi2014}: A pronounced K$\beta'$ satellite yields a high-spin state (HS), a less pronounced satellite, as in the case of FeS$_2$, yields a low-spin state (LS) in the valence band\cite{Peng1994, DeGroot2001, Joe2016}. This observation has already been made and discussed in detail by Rueff et al.\cite{Rueff1999} with the focus on pressure induced spin sensitivity of the K$\beta'$ satellite in FeS. They report on a splitting between K$\beta '$ and K$\beta_{1,3}$ of 12.5~eV which confirms our result of 12(1)~eV.

The fit of the K$\beta_{1,3}$ lines allows to extract the vtc-XES around the K$\beta_{2,5}$ line from the K$\beta_{1,3}$ high-energy tail as described for example by Gallo et al.\cite{Gallo2014}. The result is depicted in figure \ref{pic:FeS_FeS2_Kbeta} (b). Two structures labeled as K$\beta_{2,5}$ and K$\beta''$ are revealed. The K$\beta_{2,5}$ feature is generally not assigned to one specific transition from one state to the core hole but a mixture of valence states. It also involves the quadrupole 3d-1s transition \cite{Mandic2009}. More recent calculations regarding binary zinc compounds indicate that a major contribution to the K$\beta_{2,5}$ line originates from 4p metal states\cite{Mortensen2017}. The energy shift of the K$\beta_{2,5}$ line has been proven to shift with oxidation state due to screening mechanisms that depend on the valence electron density on the metal ion, i.e., the smaller the oxidation state the more the K$\beta_{2,5}$ is shifted to lower energies\cite{Bergmann2009}. In the present two sample systems Fe is in the same oxidation state. Although the position of the K$\beta_{1,3}$ line in the case of FeS shifts roughly by 1.7(5)~eV due to exchange interaction, the K$\beta_{2,5}$ line has the same emission energy (7108(1)~eV) for both FeS and FeS$_2$. 
The K$\beta''$ feature is associated with cross-over transitions from ligand orbitals and, thus, is only observed with a ligand atom present\cite{Mortensen2017, Wansleben2018}. In the case of sulfur these ligand orbitals correspond to 3s and 3p orbitals. 

\begin{figure}
	\begin{center}
	\includegraphics[width=0.55\textwidth]{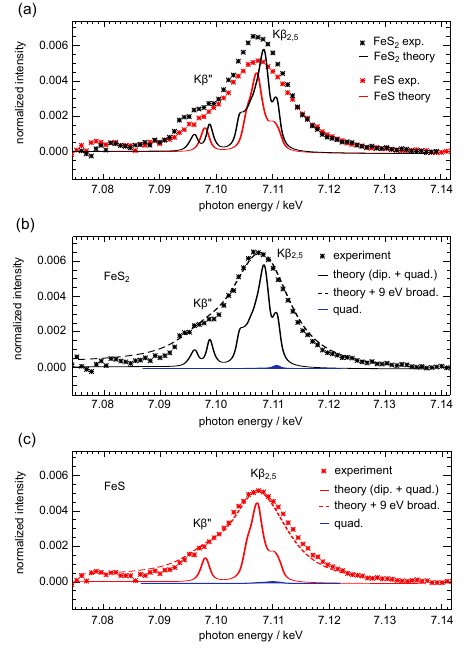}
	\end{center} 
	\caption{Experimental and calculated {\sc ocean} Fe valence-to-core spectra of FeS$_2$ and FeS. The experimental data is normalized to the total intensity of the K$\beta$ emission. The theoretical results (solid line) are shifted in energy, scaled in intensity to align with the experiment and include dipole (dip.) and quadrupole (quad.) transitions. For better comparison the theoretical results in (b) and (c) are additionally convolved with the expected spectrometer response (dashed line) which is estimated with a Lorentzian of 9~eV FWHM.}
	\label{pic:FeS2_vs_OCEAN}
\end{figure}

Figure \ref{pic:FeS2_vs_OCEAN} (a) displays the extracted experimental and calculated {\sc ocean} vtc-XES spectra of FeS and FeS$_2$. Differences are mainly seen in the amplitude of the K$\beta_{2,5}$. The K$\beta ''$ peak is also more pronounced in the case of FeS$_2$.   
The experimental data suggests in general an asymmetric structure towards the low-energy side which is not satisfactorily resolved. The asymmetry, however, indicates an ensemble of peaks. Below the experimental data, the theoretical spectra are plotted. They are both equally shifted in energy (energy offset of 7097~eV) and scaled in intensity to align with the experiment. In order to compare experiment and theory figures \ref{pic:FeS2_vs_OCEAN} (b) and (c) display the respective spectra for FeS and FeS$_2$ and a convolution of the calculated spectra with the experimental resolution achieving good agreement. The respective contributions of quadrupole transitions are included in the plots and are found to be minor. In the case of FeS$_2$ the results confirm calculations by Antonov et al.\cite{Antonov2009}. The minor contribution of the quadrupole transitions leads to the conclusion that the observed vtc-XES features originate mostly from dipole-allowed transitions between valence states with p character and the 1s vacancy. 

\section{Absorption spectroscopy}
In order to validate the capabilites of the {\sc ocean} package, we extended the comparison of experiment and theory to the Fe K NEXAFS of the two samples. Here, the unoccupied electronic states are experimentally probed which results in resonances of the absorption coefficient. 

\begin{figure}
	\begin{center}
	 \includegraphics[width=0.85\textwidth]{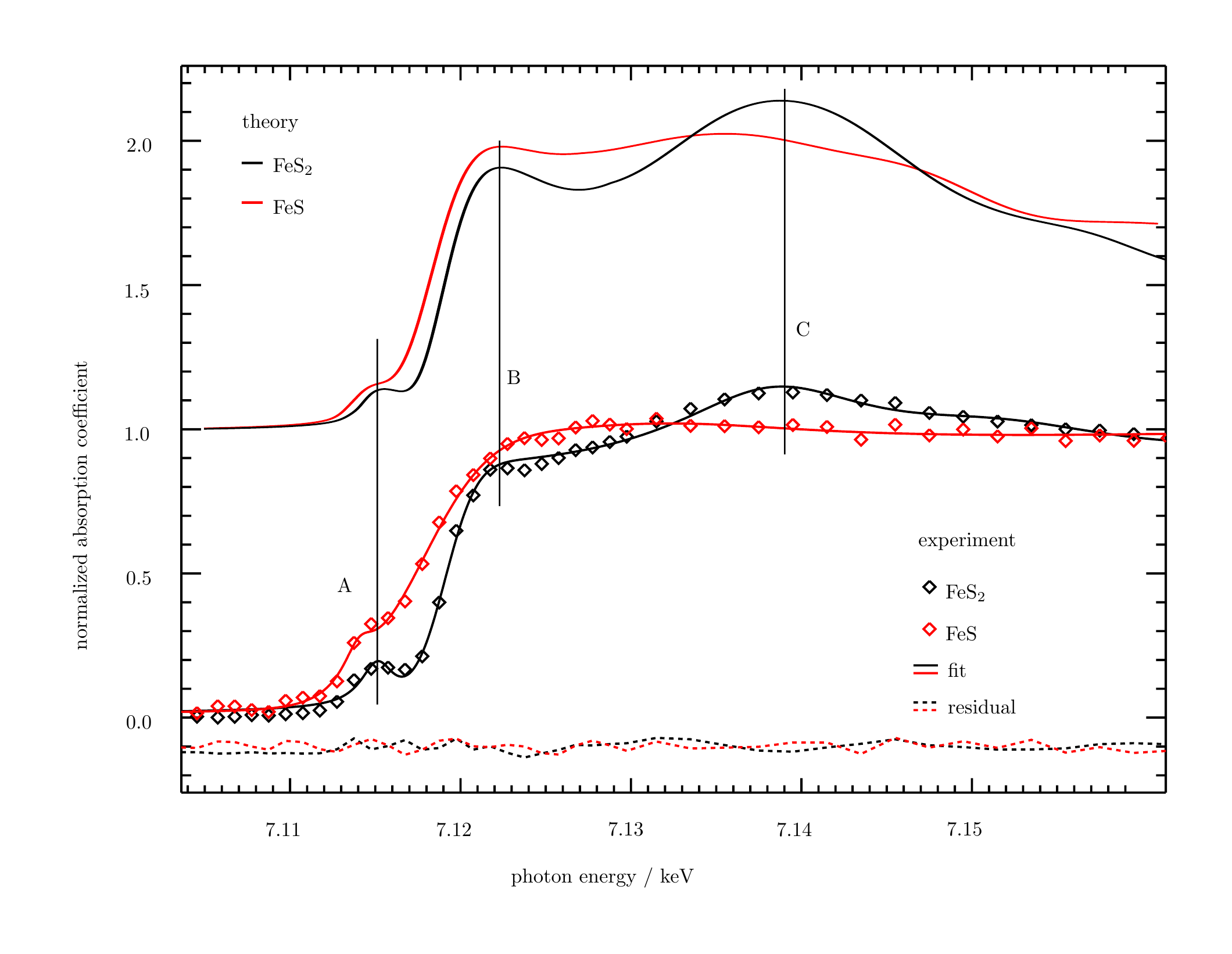}
	\end{center}
	\caption{Experimental (bottom) and calculated {\sc ocean} (top) Fe K NEXAFS of FeS$_2$ and FeS. The integration time for each photon energy was 5~seconds. The experimental data is fitted using a sum of a step function and four Gaussians. The theoretical spectra are convolved with an energy-dependent Gaussian broadening (see text) and shifted vertically for clarity. Note that the used energy offset is the same as was used for the XES calculations.}
	\label{pic:Fe_NEXAFS}
\end{figure}

The results are displayed in Figure \ref{pic:Fe_NEXAFS} where the normalized absorption coefficient is plotted as a function of incident photon energy. Experimental NEXAFS studies that explain the spectra in more detail can be found elsewhere \cite{BOSTICK2003909, KITAJOU2014391, Jahrsengene2019}. Above the experimental data, the calculated absorption spectra are plotted. The latter have been artificially broadened to better match the experimental data. A Gaussian broadening was applied with a width dependent on the absorption energy in order to account for the core hole lifetime broadening when varying the energy across the ionization threshold as well as the finite energy resolution of the incident X-ray radiation. The minimum FWHM was 0.5~eV. Starting at threshold the broadening was gradually increased  up to a maximum of 8.0~eV FWHM, a choice which was empirically guided to best match the experimental data. Indeed, it is known that the calculations do not yet account for all different sources physically contributing to a broader energy response.

Nevertheless, distinct differences in the three spectral regions labeled A, B, and C are observed between the two samples. While the pre-edge feature A of FeS is more pronounced and slightly shifted, FeS$_2$ shows an additional features C at an incident photon energy of 7.14~keV. The three labeled regions and the slight shift towards lower energy in the onset of the FeS compared to the FeS$_2$ is captured by the calculations. Note that both calculated spectra are shifted in energy by the same amount as the calculated XES data.

\section{Conclusion}
In summary, we have studied the Fe K X-ray emission and absorption spectra of FeS and FeS$_2$. Despite the moderate resolving power ($E/\Delta E = 800$) of the full-cylinder von Hamos spectrometer we demonstrate that the analysis of the chemically sensitive Fe K$\beta$ emission spectrum is still feasible. The large spectrometer band width of over 700~eV of the spectrometer enables to derive an energy scale insensitive to sample positioning by referencing the chemically less sensitive Fe K$\alpha$ emission to literature values. The occurrence of a strong K$\beta '$ satellite line in the case of FeS reveals a high valence spin state and agrees well with previously published results. The vtc-XES are extracted from the K$\beta_{1,3}$ high-energy tail and are found to be in good agreement with {\it ab initio} calculations using the {\sc ocean} software package. To validate the calculations, we used the same set of input parameters and energy offset to calculate Fe K NEXAFS spectra of the two samples and compared the results to experiment. Here, the onset of the pre-edge peak and respective resonances in the absorption coefficient match satisfactorily in energy. It could be shown that the {\sc ocean} software package is well suited for calculating XES and XAS for first row 3d transition metals. Thus, it can be used to support and validate experimental results achieved with only moderate energy resolution. Using the same approach as we demonstrated for iron sulphides, the combination of XES experiments and OCEAN calculations allows investigating the electronic structure of different transition metal sulphides for speciation purposes with sufficiently high discrimination capabilities.

\section{Appendix}
Certain commercial equipment, instruments, or materials are identified in this paper in order to specify the experimental procedure adequately. Such identification is not intended to imply recommendation or endorsement by the National Institute of Standards and Technology, nor is it intended to imply that the materials or equipment identified are necessarily the best available for the purpose.

\section{Acknowledgements}
Parts of this research was performed within the EMPIR project Advent. This project has received funding from the EMPIR programme co-financed by the Participating States and from the European Union\' s Horizon 2020 research and innovation programme.

%-------------------------------------------------------------------------------------------------------
%-------------------------------------------------------------------------------------------------------
\cleardoublepage
\bibliography{rsc}

\end{document}